\documentclass[letterpaper%
, oneside%
, 12pt%
,these%
, english%
,creativecommons, withAlgo2e %
]{thETS}

\usepackage{float}
\usepackage{listings}
\usepackage{subfig}
\usepackage{hyphenat}
\usepackage{color,soul}

\newcites{refs}{LIST OF REFERENCES}

\title{Channel Modeling for Over-water Communications}

\author{Ahmed Abdelmoaty}
\authorcopyright{Ahmed Abdelmoaty}

\datesoutenance{"Defense date"}

\datedepot{"December, 2018"}

\directeur{Prof.}{Francois Gagnon}{École de technologie supérieure}

\codirecteur{Dr.}{Ghassan Dahman}{Ultra Electronics}

\begin{document}

\pagenumbering{Roman}

\maketitle

\presentjury

%
%








\begin{abstract}{Ray Tracing, Parabolic Equations, Over-Water Propagation}

Over-water communication has been an important research field of wireless communications and radar. Since there are so many parameters in this environment that are changing continuously in the spatial and temporal domains such as  weather and sea surface parameters. Hence, accurately modeling the over-water propagation channel is very crucial and challenging at the same time.
In this report, we will shed some light on the parabolic equation (PE) method which is one of the main methods used to model the radio waves propagation over variable terrain and through homogeneous and inhomogeneous atmosphere. Besides ray-tracing, PE is among the most widely used approaches nowadays for deterministically analyze over-water channels. We are trying to pave the way for better understanding for the mathematical background for the PE and develop a tool to perform advanced analysis for the over-water communication channel taking into consideration its spatial and temporal behavior.


\end{abstract}

\tableofcontents


\listoffigures


\begin{listofabbr}[3cm]
\item [2W-SSPE] Two Way Split-Step Parabolic Equations
\item [AREPS] Advanced Refractive Effects Prediction System
\item [ASF] Additional Secondary Factor
\item [BS] Boundary Shift
\item [DEM] Digital Elevation Method
\item [DMFT] Discrete Mixed Fourier Transform
\item [FDM] Finite Difference Method
\item [FEM] Finite Element Method
\item [GIS] Geographic Imaging System
\item [GTD] Geometric Theory of Diffraction 
\item [GUI] Graphical User Interface 
\item [HO-FDTD] High-Order Finite Difference Time Domain
\item [IMC] Impedance Boundary Conditions
\item [IE] Integral Equation
\item [ITU] International Telecommunication Union
\item [MFT] Mixed Fourier Transform
\item [MoM] Method of Moment
\item [NAPE] Narrow-Angle Parabolic Equations
\item [ODE] Ordinary Differential Equations
\item [PE] Parabolic Equations
\item [PEC] Perfectly Electrical Conducting
\item [PML] Perfect Matching Layer
\item [RT] Ray Tracing
\item [SBR] Shooting and Bouncing Ray
\item [SRTM] Shtttle Radar Topography Mission
\item [SSFT] Split-Step Fourier Transform
\item [SSPE] Split-Step Parabolic Equations
\item [SSVM] Split-Step Wavelet Method
\item [VV$\&$C] Validation, Verification and Calibration
\item [WAPE] Wide-Angle Parabolic Equations
\end{listofabbr}


\cleardoublepage

\pagenumbering{arabic}

\reversemarginpar


\chapter{Introduction}
Wireless communications occupies a paramount role in our daily activities, due to the very involved wireless applications such as such as: Mobile phones, Wi-Fi, controlling remote environments, Satellite communications (including satellite TV channels broadcasting),…etc \cite{afifi2016telepresence,korashy2016teleoperation,taj2016new}.     
One of the most crucial parts of any communication system is the wireless channel. The wireless channel get its importance from the fact that it is the medium in which the signal or information is propagated. Although electromagnetic waves are fully characterized by Maxwell equations, in realistic propagation scenarios it is difficult to have an analytical solution. Therefore, one way to simplify this problem is to approach it using deterministic approach. The main objective of deterministic propagation modeling is to have an approximate description of the propagation scenario such as antenna heights, terrain descriptions, frequency of operation, etc, and then an appropriate approach can be used to solve for the field strength. In the tropospheric propagation, the inhomogeneous tropospheric refractivity profile resulting from variable weather and sea-surface conditions is the most efficient way to describe the propagation medium behavior. \cite{farah2016new,farah2017new}.   
 
\begin{figure}[H]
\captionsetup{justification=centering}
\centering
\includegraphics[width=15cm]{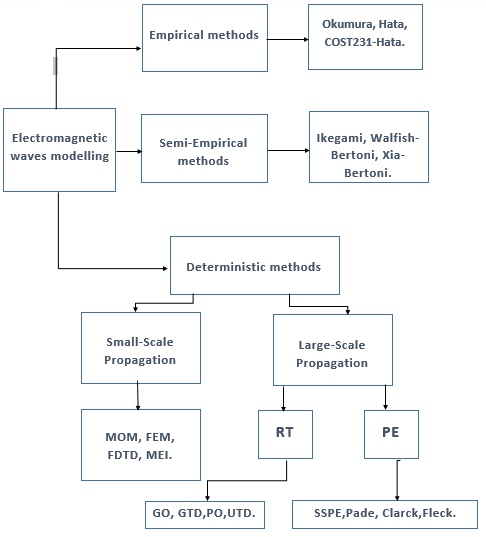}
\caption{Illustration of propagation models.}
\label{EM tree}
\end{figure}

From the early and the simplest propagation models the one which was introduced by Friis for free space radio propagation in 1946 \cite{yun2015ray}, the continuous efforts of the scientists and researchers to develop this model and built on it did not stop. Propagation models vary between: empirical models, semi-empirical/semi-deterministic models, and deterministic models see Fig. \ref{EM tree}. The empirical models (i.e., Okumura, Hata…etc.) are the simplest ones and can only give statistical results which are only valid on average. The semi-empirical models or semi-deterministic models are limited for small-scale application such as indoor, and microcell. The deterministic models grasp the attention of the researches for many years, where it was built on approximate solutions of Maxwell equation and can be applied for a wide range of applications. Moreover, the accuracy of the solution, and the amount of memory occupied are not comparable with other methods \cite{akbarpour2005ray,shen2012comparisons}. 

Specifically, the Ray Tracing and the Parabolic Equation methods are the deterministic models which are suitable for long range large–scale propagation; we mean by large-scale here the propagation with the presence of  complex environments such as irregular terrain, and different atmospheric conditions, due to their computational efficiency and accuracy. Interestingly, the deterministic models may be randomized by using the statistical distribution of atmospheric conditions. If the deterministic model is precise enough, the resulting statistical model should permit to accurately predict the quality of the communication link, for example the average throughput and outage probabilities. Proceeding from this brief introduction we are going to discuss the two last methods in some detailed fashion.

\section{Radio Propagation Modeling using Ray Tracing}
The concept of rays is a very logical and intuitive concept which can be experienced in everyday of our life. Simply, the sun light which enters your window every morning is a very direct example of “ray” which propagates directly from the sun to your room’s window. For radio wave propagation, the ray concept is well established by the high frequency approximation of Maxwell’s equation. Using these equations with the assumption of high frequencies, and accompanied by the help of Ampere’s law, Gauss’s law, and Faraday’s law, for the electric and magnetic fields, the reflection, refraction, and diffraction of rays can be calculated \cite{shen2012comparisons,dahman2017ship}.

In the light of the Fermat’s principle of least time which states that the ray will take a route which consumes the least time possible to travel from on point to another, the ray concept to serve the propagation using ray tracing can be summed as:      
\begin{itemize}
\item  For homogeneous medium, the ray travels in a straight line. 
\item  The laws of diffraction, refraction, and reflection are applied.
\item  The ray’s energy is contained and propagated in a tube (surrounding this central ray).
\end{itemize}

\begin{figure}[H]
\captionsetup{justification=centering}
\centering
\includegraphics[width=15cm]{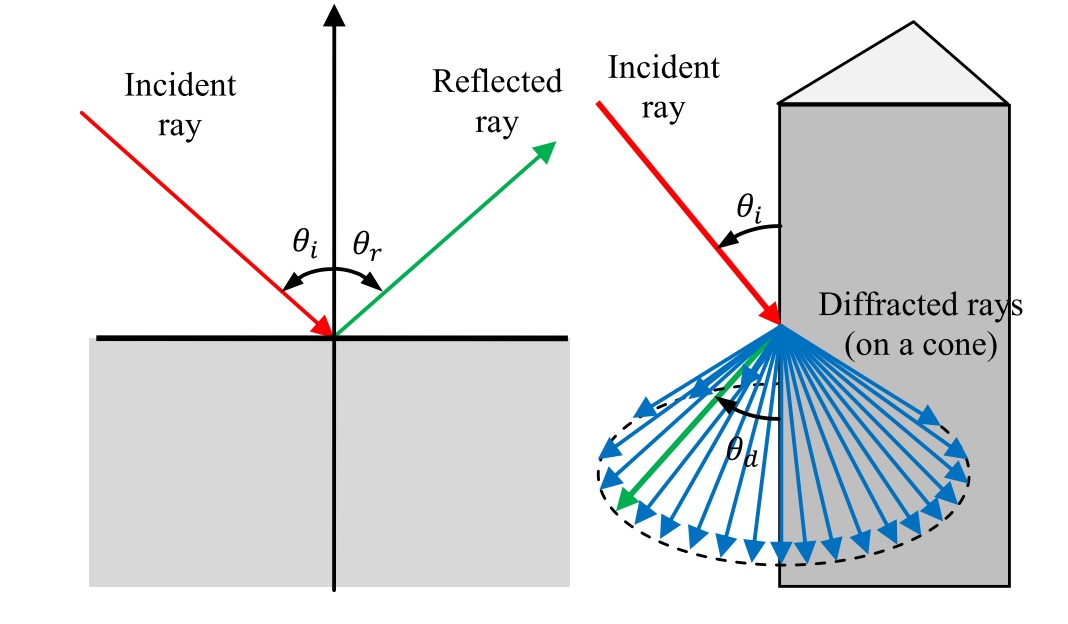}
\caption{Different ray types.}
\label{raytypes}
\end{figure}

There are many types of rays, direct rays, diffracted rays and reflected rays. Fig. \ref{raytypes} gives an illustration for the different types of rays.  Many algorithms are used to find the rays' trajectories as the wave propagates, for example, Fermat’s principle of least time, the image method, the shooting and bouncing Ray (SBR) method, and different hybrid methods. 

\section{Radio Propagation Modeling using Parabolic Equation}
\begin{figure}[H]
\captionsetup{justification=centering}
\centering
\includegraphics[width=12cm]{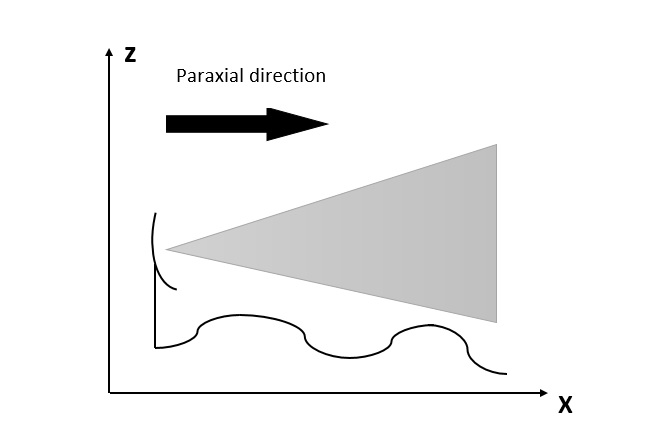}
\caption{A schematic of a tropospheric propagation.}
\label{schematic}
\end{figure}

In all that follows, we assume $\exp{(-i\omega t)}$ time-dependence of the fields, where $\omega$ is the angular frequency. Initially we work with Cartesian coordinates $(x,y,z)$. In this report, we are concerned with two-dimensional electromagnetic problems where the fields are independent of the transverse coordinate y. There are then no depolarization effects, and all the fields can be decomposed into horizontally and vertically polarized components propagating independently. For horizontal polarization, the electric field $E$ only has one non-zero component $E_y$, while for vertical polarization, the magnetic field $H$ only has one non-zero component $H_y$. We work with the appropriate field component $\psi$ defined by, for horizontal polarization:
\begin{align}
&\psi (x,z) = E_{y} (x,z)
\end{align}
and for vertical polarization
\begin{align}
&\psi (x,z) = H_{y} (x,z)
\end{align}
We are interested in solving problems where energy propagates at small angles from a preferred direction, called the paraxial direction. Following the convention in radiowave propagation problems, we choose the positive x-direction as the paraxial direction. Fig. \ref{schematic} gives an example for the tropospheric propagation.\\
If the propagation medium is homogeneous with refractive index $n$, the field component $\psi$ satisfies the two-dimensional scalar wave equation
\begin{align}
\frac{\partial^2 \psi}{\partial x^2} + \frac{\partial^2 \psi}{\partial z^2} + k^2 n^2 \psi = 0
\end{align}
where $k$ is the wave number in vacuum.
\subsection{Paraxial wave equation}
The reduced function associated with the paraxial direction $x$ is given as
\begin{align}
u(x,z) = {\rm e}^{-ikx} \psi (x,z)
\end{align}
the reduced form of the function is useful since it is slowly varying in range with angels close to the paraxial direction of propagation.\\
The scalar wave equation in terms of $n$ is given as
\begin{align}
\frac{\partial^2 u}{\partial x^2} + 2ik \frac{\partial u}{\partial x}+\frac{\partial^2 u}{\partial z^2} + k^2 \left( n^2 -1 \right) u = 0
\end{align}
 which can be factorized as 
\begin{align}\label{Op}
\biggl\{ \frac{\partial}{\partial x} + ik\left( 1-Q\right)\biggr\} \biggl\{ \frac{\partial}{\partial x} + ik\left( 1+Q\right) \biggr\}u = 0
\end{align}
Where $Q$ is known as a pseudo-differential operator and given as 
\begin{align}
Q = \sqrt{\frac{1}{k^2} \frac{\partial^2}{\partial z^2} + n^2 (x,z)}
\end{align}
To continue with our framework, we need to give a suitable mathematical frame work for the square root symbol at the expression of $Q$. The square root can be seen as a composition of operators such as:
\begin{align}
Q (Q(u)) = {\frac{1}{k^2} \frac{\partial^2}{\partial z^2} + n^2 (x,z)}
\end{align}
Now, we need to split the wave equation into the terms which are given in Eq. (\ref{Op}) and try to find a function which satisfying the pseudo-differential equations
\begin{align} \label{frw}
\frac{\partial u}{\partial x}  = -ik(1 - Q)u
\end{align}  
\begin{align} \label{bkw}
\frac{\partial u}{\partial x}  = -ik(1 + Q)u
\end{align}
Where Eqs. (\ref{frw}) \& (\ref{bkw}) corresponds to forward and backward propagation respectively.\\
Since the solution of Eq.(\ref{frw}) neglects the backscattered field, it's not reflect the actual electromagnetic field. However, these solutions are accurate enough for range-independent medium where there is no need for the factorization of Eq.(\ref{Op}). In order to have an exact solutions for the reduced function, Eqs.(\ref{frw}) \& (\ref{bkw}) should be solved simultaneously in the form
\begin{align} 
 \left\{ \begin{array}{lr} 
 u  &  \text {=} \qquad u_+ + u_- \\
 \frac{\partial u_+}{\partial x}  &  \text {=} \qquad -ik(1 - Q)u_+ \\
 \frac{\partial u_-}{\partial x}  &  \text {=} \qquad -ik(1 + Q)u_-
\end{array}
\right.
\end{align}   
Its clear that Eqs. (\ref{frw}) \& (\ref{bkw}) are pseudo-differential equations of first order which can by solved by marching techniques. In other words, the solution at a certain range can be simply obtained by knowing the initial field and the top and down boundary conditions of the domain. For example the solution of the forward propagation eq.(\ref{frw}) is given as
\begin{align}
u(x+\Delta x , .) = {\rm e}^{ik\Delta x(-1+Q)} u (x ,.)
\end{align}    
\section{Vector Parabolic Equation}
The scalar PE method is typically applied with PEC surfaces, where the transverse magnetic and electric fields propagate independently. However, this is not always the case for the practical work where there are lossy surfaces and the coupling between the TE and TM modes (depolarization effect) cannot be neglected. Hence, using the scalar PE cannot be used anymore. Consequently, VPE side by side with Impedance Boundary Conditions (IBC) was introduced to deal with this type of problems.
The standard VPE was formulated by as \cite{popov2000modeling}:
\begin{align}
\frac{\partial T}{\partial x} = \frac{1}{2jk} \left( \frac{{\partial}^2}{\partial y^2} + \frac{{\partial}^2}{\partial z^2} \right) T
\end{align} 
Where $T$ is the complex wave amplitude.

\section{Comparison between RT and PE}

\begin{figure}[H]
\captionsetup{justification=centering}
\centering
\includegraphics[width=15cm]{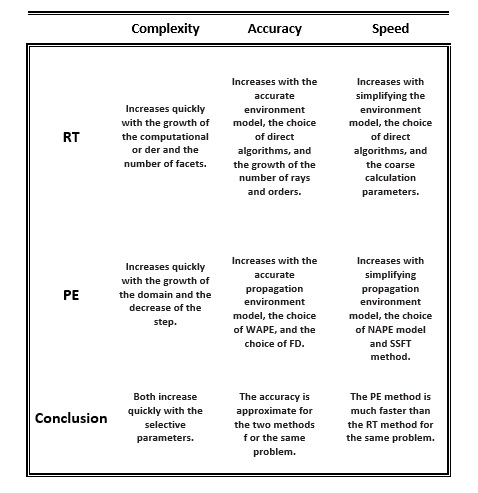}
\caption{Comparison of RT and PE methods.}
\label{Comparison}
\end{figure}

A comparison between the PE and RT in terms of accuracy, speed of calculation, and the complexity of solution models was introduced in \cite{shen2012comparisons}. the comparison can be summarized in Fig. \ref{Comparison}.   





\section{Hybrid Models}
From the above discussion, we may conclude that, ray-tracing methods enable propagation modeling in very complicated scenarios and they can provide reasonably accurate prediction of signal fading characteristics. However, these methods typically have an exponential computational complexity for certain practical problems especially when the reflection level and the separation between the transmitter and receiver are increased. On the other hand, the PE methods are more computationally efficient and since they are built on the famous Helmholtz equation they are count for the diffraction and refraction effects very well. However, Vectorial PE (VPE) methods cannot model high-order modes and corresponding rapid fluctuations accurately. Moreover, practical antenna patterns cannot be readily incorporated as the initial conditions for VPE \cite{martelly2009adi}.\\
Consequently, the needs for hybrid methods which avoids the cons of the previous methods are raised. 

In \cite{martelly2009adi}, the authors proposed a new hybrid model for predicting the propagation in tunnels. The new model mixing between the VPE and the Alternate Direction Implicit (ADI) method.
 The ADI was introduced to reduce the computational burden of VPE especially when the number of meshes increased. ADI is considered to be a better replacement for the Crank-Nicolson marching technique in PE. The proposed model was tested with number of waveguides i.e., rectangular and circular with the appropriate boundary conditions. Moreover, it was tested against empirical rectangular, circular models of tunnels, and against real data collected from TELICE Laboratory, University of Lille, France, for the Japanese railway tunnel. The results show that the proposed ADI technique is better than the conventional Crank-Nicolson in terms of execution time.     
  
Zhang et al. conducted a number of research about hybrid models for modeling train communication channels \cite{zhang2014hybrid,zhang2015efficient,zhang2016hybrid}. The hybrid model here basically between RT and VPE. The main idea is to divide the whole domain of the propagation into two sub-domains and coupled them with an appropriate interface. The first domain which includes a complex environment such as train stations can be modeled using TR. Hence, the VPE can be used to model the long range propagation through the train tunnel. The objective of the interface is to deliver the fields between the two sub-domain solvers. The question of the searching for the perfect position of the interface was solved by using the basic assumption of the VPE propagation which states that the propagation is in paraxial direction with angles up to 15, such that the received fields were divided into categories the first which is 0-15 degree and the second 15-90 degree, then the power densities are computed and weighted. The interface is positioned in a plane with power densities of rays larger than 15 degree can be neglected.  The hybrid model was tested in both environment i.e., train station on tunnel and an open air guide-way. The simulation results of the hybrid model compared to the on-site data show that the hybrid model can give almost the same results when the interface located at the right position. \\
Where the hybrid models took its place in some applications such as train and tunnel communications, the area of oversea communication still lack some investigations for the hybrid models which highlighted our recent interest of this topic.   

\chapter{Related Work}

 \section{Algorithms of PE}
\begin{figure}[H]
 \captionsetup{justification=centering}
 \centering
 \includegraphics[width=12cm]{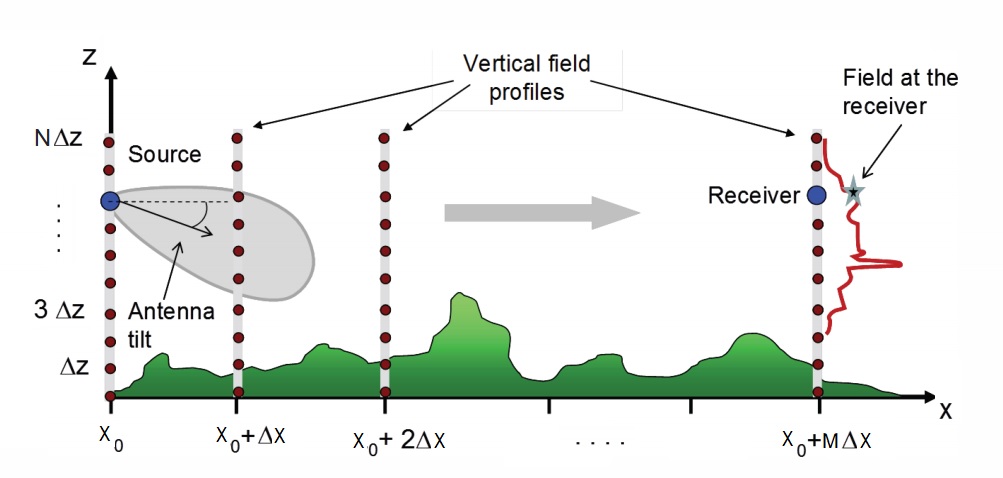}
 \caption{SSPE and flat earth implementation.}
 \label{SSPE-source}
 \end{figure}

The most commonly used algorithms for PE solution are Split-Step Fourier Transform (SSFT), Finite Difference Method (FDM), and Finite Element Method (FEM). SSFT is considered to be an analytical method, while FDM, FEM, and other such finite methods are numerical methods. The main aspect of choosing one of these algorithms is the specification of the problem. For example, in long-range propagation, SSFT is suitable for solving PEs because of its high calculation speed, numerical stability, and small calculation memory. While Finite methods are the best choice for short-range propagation because of their capability and flexibility to handle complex boundary conditions.\\
The split step fast Fourier transform (SSFT)-based PE solution uses the marching through the paraxial direction technique, see Fig. \ref{SSPE-source}. First, the initial height profile is represented by the means of an antenna pattern. The initial field is then propagated longitudinally from $X_0$ to $X_0 +\Delta X$ , and the transverse field profile at the next range is obtained. This new height profile is then used as the initial profile for the next step, the process is repeated until the propagator reaches the desired range. the mixed Fourier transform MFT is used to model The ground losses with the bottom boundary conditions. For the artificial reflections, it can be removed by extending the maximum height, then smoothly reducing the field in the extended region. Usually, windowing functions (such as Hanning, Hamming) may be applied to eliminate these reflections.
In \cite{apaydin2010numerical} a novel FEM-based surface wave multi-mixed path propagator is developed. the idea behind FEM method in solving PE is to partition the domain into sub-domains or elements. In vertical domain, the field is approximated in selected nodes and the propagation is continue in the paraxial direction using the marching technique. The authors introduced a novel propagator for implementing their algorithm called (FEMIX). The model was verified by comparing the results with the Millington curve fitting method and ITU-R P-368 curves, a great matching was found.\\
Again, the same authors conducted a detailed comparison between SSFT-based and FEM-based solution for PE \cite{apaydin2010split}. A novel MATLAB-based propagation tool which called LINPE was developed. The tests and comparisons of the SSFT and the FEM parabolic equation were done using the LINPE with comparison with an analytical solution. The result showed that FEM is faster and occupies less memory than SSFT when using the same step length. However, the accuracy of FEM is not as good as that of SSFT.  
 
 \subsection{New Algorithms for PE solution}
In the past few years, the efforts of the researchers continued to find new algorithms for solving the PE problem. Besides the aforementioned three methods. In 2011, the first attempt to use the split-step wavelet method  (SSWM) to solve PE problems was proposed \cite{iqbal2011split}. The PE problem was reduced to Ordinary Differential Equation (ODE) problem by using Galerkin projection method with periodic Daubechies scaling function. The main objective of this formulation is to discretize the height operator. The ODE problem was solved using a marching technique after tacking the effects of diffraction and refraction separately. To verify their model, The SSWM results was compared with AREPS. The outputs show that the proposed algorithm is nearly as good as AREPS which make it a good alternative to SSPE. In 2012, the same authors of \cite{iqbal2011split} proposed a novel method based on Split Step Wavelet for horizontally inhomogeneous environments \cite{iqbal2012numerical}. Unlike their last research, they consider the irregular and inhomogeneous environments by modeling the bottom boundary conditions i.e., Dirichlet, Neumann. Furthermore, the top boundaries were modeled using Perfectly Matching Layer (PML) and windowing functions for avoiding the artificial reflections. To come up with a verified model, they compare their results to the SSPE results under the same environmental conditions. A great agreement has been obtained. However, the SSWM method has large meshing requirements that take up a considerable amount of memory. From another perspective, in \cite{zhao2013new}, a new method for modeling PE was introduced. The authors proposed integrating the convolution theorem with the Fourier transform for transferring the PE from second order PDE to constant coefficient first order differential equation in the spectral domain. the objective of the new model was to improve the calculation errors caused by ignoring changes in the refractive index attributed to increases in lateral distance. The new model was validated with designed analytical solution and the well-known SSPE solution. Although the result was satisfactory, there were two main problems with the proposed model. Firstly, the assumption of smooth and perfectly conduction bottom boundary, which well introduce a difficulty when it comes to a complex terrain surface. Secondly, the computational burden and the stability issues that are arises from the coefficient matrix.
 \section{Irregular terrain modeling}
 In \cite{guan2018parabolic}, the authors proposed the Digital Elevation Method (DEM) over the parabolic equation forward propagation to predict the propagation over irregular terrain area. They obtained the wide angle parabolic equation and used the split step Fourier transform algorithm to solve it. They validate the proposed model by comparing it to the shift-map (SP) method over a perfectly conducting Gaussian terrain profile. They noticed that their model is perfectly matches the shift map method results unless the resolution of the DEM is exceeds the range step of the PE method.\\
 A comparative study between the two PE approaches, narrow angle PE (NAPE) and wide angle PE (WAPE) and High-Order Finite Difference Time Domain (HO-FDTD) method was introduced in \cite{paradacomparative}. The main objective of the study was to compare the performance of the SSPE versus HO-FDTD over irregular terrain and lossy environment.  The irregular terrain was initially approximated by staircase approximation, where the imperfect conducting surfaces were modeled by using Discrete Mixed Fourier Transform (DMFT). In terms of accuracy, the two approaches of SSPE matching the results obtained by HO-FDTD, while the SSPE approaches outfit the HO-FDTD in terms of time of computation. \\
In \cite{5738678}, the authors proposed a new algorithm for inhomogeneous earth terrain modeling. The new algorithm exploits PE with Finite Element Method (FEM) to model forward and backward propagation. Basically, the two way FEMPE divides the transverse domain which its boundaries defined between the ground and the user-defined maximum height into a number of elements. Then, starting from the initial field, the approximated field values at the selected discrete nodes in the vertical domain are propagated at the paraxial direction. The authors verified the validity of their model by setting a comparison with the exact analytical solution for certain propagation model such as parallel plate with Perfect Electrically Conducting (PEC) boundaries, and with the split step Fourier transform PE model. The results show that they perfectly matches the SSPE results in different atmospheric and terrain scenarios. The only constraint on FEMPE is that the steps size of both vertical and horizontal direction should be small to avoid the numerical oscillation problem. As a result of this constraint the required time for solving the FEMPE problem is much larger (14 times) of the SSPE problem. On the other hand, the FEMPE algorithm is outfitting the SSPE in modeling different kinds of boundary conditions.\\
The authors of \cite{bai2013prediction}, proposed a new method for predicting the terrain profile for wave propagation application using the Shuttle Radar Topography Mission (SRTM) data. SRTM uses a bilinear interpolation technique to get the elevation value of any arbitrary location at the terrain profile. Then, the PE method is used to predict the wave propagation behavior based on the predicted terrain profile. The elevation data usually stored in a grid format. To obtain the elevation value of a certain sampling interval, an interpolation method is used to get that done. In this paper the authors used the bilinear interpolation method. The terrain profile predicted by the proposed method was compared with the one extracted by Geographic Imaging System (GIS) to validate its effectiveness. The results show that there were a perfect match between the terrain models predicted by SRTM and GIS. The benefit of using this technique for predicting the terrain profile is its quickness and accuracy.\\
The problem of varying terrain was tackled in \cite{holm2007wide}. The authors proposed two approaches for dealing with this problem. The first one called the first order solution which based on the coordinate transformation to incorporate the terrain variation, this technique was introduced by Donohue and Kuttler in \cite{donohue2000propagation} and known as the shift map technique and its suitable for narrow angle PE problems. The second approach is the second order scheme by using Pade approximation and it is used with the wide angle PE problems. In both techniques the piecewise linear approximation for the terrain was used. To validate the two schemes, the authors compared the two PE based schemes with the Geometric Theory of Diffraction (GTD) results for varying terrain.  The proposed techniques of PE matching the results of GTD results with good accuracy. The PE techniques have the advantage over the GTD that GTD is more time consuming when the number of wedges is increased. \\
The earliest version of the MATLAB based GUI of parabolic equation model for wave propagation was first proposed in \cite{sevgi2005matlab}. The PE model here considers only the forward propagation. The algorithm which used for solution is the split step Fourier transform such that the GUI called SSPE-GUI. The recent two way PE model can be considered as an extension for the work of those authors.\\
Most of the researchers dealt the rooftops of the irregular terrain as a PEC material, which is not practical assumption. In \cite{wang2012propagation}, the authors proposed the improved 2W-SSPE which incorporates the modeling of the rooftops of the building as lossy surfaces or even a dielectric materials. The proposed model uses DMFT with Boundary shift (BS) to model the impedance boundary conditions for irregular terrain.{(You may need to add something about boundary shift)}. The results of the proposed work were shown in terms of a comparison between the conventional 2W-SSPE and the improved 2W-SSPE. The results show that the conventional 2W-SSPE is more dependent in the electrical nature of the irregular surfaces. On the other hand, the improved 2W-SSPE is more flexible and more accurate in modeling different types of complex surfaces which tends to make it more practical.\\
The problem of time delay due to propagation over irregular paths and varying electric surfaces named Additional Secondary Factor (ASF). The ASF problem is crucial for long wave navigation system, where the small deviation in predicting the ASF correctly may results in a huge errors up to a number of Kilometers. The authors in \cite{wang2016parabolic}, introduced a new model for predicting the ASF based on SSPE method and compare it to the two well-known methods for prediction the ASF namely Integral Equation (IE) and Finite Difference Time Domain (FDTD) methods. Basically, The IE method is based on the assumption that the impacts from backscattering waves are very small and can be neglected. As expected, it has a large error for steeper terrain and choppy electric parameters and is hard to be used for a three-dimensional (3-D) case because of the vulnerability of the algorithm itself. The FDTD method has been proven to be most the precise one for complex paths with irregular terrain, but the computational expenditures of memory and time are huge for the large area prediction. The results show that the proposed model of Loran-C ASP prediction based on SSPE is matching the results of the IE and FDTD in terms of accuracy, but it outfits them in terms of computational time by several orders.     
\section{Initial field modeling}
Initial field modeling or source modeling is considered to be one of the most important parts of PE algorithms. Usually, the researchers In long-range propagation models the source as Gaussian antenna beam, since it is easily to adjust beamwidth and beam tilt and provide a good representation for paraboloid dish antennas. In short-range propagation, excitation differences cannot be neglected in source modeling. A detailed comparative study has been conducted by the authors of \cite{apaydin2013groundwave} on accurate source modeling. They performed a very detailed processes of Validation, Verification, and Calibration VV$\&$C for the most widely used analytical source modeling methods such as Gaussian beam pattern and line-source, compared to the method of moments MoM, FEM-PE, SSFT-PE.\\

\newpage
\begin{spacing}{1}
%


\begin{thebibliography}{10}
\providecommand{\url}[1]{#1}
\csname url@samestyle\endcsname
\providecommand{\newblock}{\relax}
\providecommand{\bibinfo}[2]{#2}
\providecommand{\BIBentrySTDinterwordspacing}{\spaceskip=0pt\relax}
\providecommand{\BIBentryALTinterwordstretchfactor}{4}
\providecommand{\BIBentryALTinterwordspacing}{\spaceskip=\fontdimen2\font plus
\BIBentryALTinterwordstretchfactor\fontdimen3\font minus
  \fontdimen4\font\relax}
\providecommand{\BIBforeignlanguage}[2]{{%
\expandafter\ifx\csname l@#1\endcsname\relax
\typeout{** WARNING: IEEEtranS.bst: No hyphenation pattern has been}%
\typeout{** loaded for the language `#1'. Using the pattern for}%
\typeout{** the default language instead.}%
\else
\language=\csname l@#1\endcsname
\fi
#2}}
\providecommand{\BIBdecl}{\relax}
\BIBdecl

\bibitem{afifi2016telepresence}
M.~Afifi, M.~Korashy, A.~H. Ahmed, Z.~Hafez, and M.~Nasser, ``Telepresence
  robot using microsoft kinect sensor and video glasses,'' in \emph{The 1st
  International Conference on Advanced Intelligent System and Informatics
  (AISI2015), November 28-30, 2015, Beni Suef, Egypt}.\hskip 1em plus 0.5em
  minus 0.4em\relax Springer, 2016, pp. 91--101.

\bibitem{akbarpour2005ray}
R.~Akbarpour and A.~R. Webster, ``Ray-tracing and parabolic equation methods in
  the modeling of a tropospheric microwave link,'' \emph{IEEE Transactions on
  Antennas and Propagation}, vol.~53, no.~11, pp. 3785--3791, 2005.

\bibitem{5738678}
G.~Apaydin, O.~Ozgun, M.~Kuzuoglu, and L.~Sevgi, ``A novel two-way
  finite-element parabolic equation groundwave propagation tool: Tests with
  canonical structures and calibration,'' \emph{IEEE Transactions on Geoscience
  and Remote Sensing}, vol.~49, no.~8, pp. 2887--2899, Aug 2011.

\bibitem{apaydin2010numerical}
G.~Apaydin and L.~Sevgi, ``Numerical investigations of and path loss
  predictions for surface wave propagation over sea paths including hilly
  island transitions,'' \emph{IEEE Transactions on Antennas and Propagation},
  vol.~58, no.~4, pp. 1302--1314, 2010.

\bibitem{apaydin2010split}
G.~Apaydin and L.~Sevgi, ``The split-step-fourier and finite-element-based
  parabolic-equation propagation-prediction tools: Canonical tests, systematic
  comparisons, and calibration,'' \emph{IEEE Antennas and Propagation
  Magazine}, vol.~52, no.~3, pp. 66--79, 2010.

\bibitem{apaydin2013groundwave}
G.~Apaydin and L.~Sevgi, ``Groundwave propagation at short ranges and accurate
  source modeling [testing ourselves],'' \emph{IEEE Antennas and Propagation
  Magazine}, vol.~55, no.~3, pp. 244--262, 2013.

\bibitem{bai2013prediction}
R.~Bai, C.~Liao, N.~Sheng, and Q.~Zhang, ``Prediction of wave propagation over
  digital terrain by parabolic equation model,'' in \emph{Microwave, Antenna,
  Propagation and EMC Technologies for Wireless Communications (MAPE), 2013
  IEEE 5th International Symposium on}.\hskip 1em plus 0.5em minus 0.4em\relax
  IEEE, 2013, pp. 458--461.

\bibitem{dahman2017ship}
G.~Dahman, F.~Gagnon, and G.~Poitau, ``Ship-to-ship beyond line-of-sight
  communications: a comparison between ray tracing simulations and the
  petool,'' in \emph{General Assembly and Scientific Symposium of the
  International Union of Radio Science (URSI GASS), 2017 XXXIInd}.\hskip 1em
  plus 0.5em minus 0.4em\relax IEEE, 2017, pp. 1--4.

\bibitem{donohue2000propagation}
D.~J. Donohue and J.~Kuttler, ``Propagation modeling over terrain using the
  parabolic wave equation,'' \emph{IEEE Transactions on Antennas and
  Propagation}, vol.~48, no.~2, pp. 260--277, 2000.

\bibitem{farah2016new}
J.~Farah, E.~Sfeir, C.~A. Nour, and C.~Douillard, ``New efficient energy-saving
  techniques for resource allocation in downlink ofdma transmission systems,''
  in \emph{Computers and Communications (ISCC), 2017 IEEE Symposium on}.\hskip
  1em plus 0.5em minus 0.4em\relax IEEE, 2017, pp. 1056--1062.

\bibitem{farah2017new}
J.~Farah, E.~Sfeir, C.~A. Nour, and C.~Douillard, ``New resource allocation
  techniques for base station power reduction in orthogonal and non-orthogonal
  multiplexing systems,'' in \emph{Communications Workshops (ICC Workshops),
  2017 IEEE International Conference on}.\hskip 1em plus 0.5em minus
  0.4em\relax IEEE, 2017, pp. 618--624.

\bibitem{guan2018parabolic}
X.-W. Guan, L.-X. Guo, Y.-J. Wang, and Q.-L. Li, ``Parabolic equation modeling
  of propagation over terrain using digital elevation model,''
  \emph{International Journal of Antennas and Propagation}, vol. 2018, pp.
  1--6, 2018.

\bibitem{holm2007wide}
P.~D. Holm, ``Wide-angle shift-map pe for a piecewise linear terrain—a
  finite-difference approach,'' \emph{IEEE Transactions on Antennas and
  Propagation}, vol.~55, no.~10, pp. 2773--2789, 2007.

\bibitem{iqbal2011split}
A.~Iqbal and V.~Jeoti, ``A split step wavelet method for radiowave propagation
  modelling in tropospheric ducts,'' in \emph{RF and Microwave Conference
  (RFM), 2011 IEEE International}.\hskip 1em plus 0.5em minus 0.4em\relax IEEE,
  2011, pp. 67--70.

\bibitem{iqbal2012numerical}
A.~Iqbal and V.~Jeoti, ``Numerical modeling of radio wave propagation in
  horizontally inhomogeneous environment using split-step wavelet method,'' in
  \emph{Intelligent and Advanced Systems (ICIAS), 2012 4th International
  Conference on}.\hskip 1em plus 0.5em minus 0.4em\relax IEEE, 2012, pp.
  200--205.

\bibitem{korashy2016teleoperation}
M.~Korashy, K.~F. Hussain, and H.~Ibrahim, ``Teleoperation of dogs using
  controlled laser beam,'' in \emph{Digital Information and Communication
  Technology and its Applications (DICTAP), 2016 Sixth International Conference
  on}.\hskip 1em plus 0.5em minus 0.4em\relax IEEE, 2016, pp. 45--49.

\bibitem{martelly2009adi}
R.~Martelly and R.~Janaswamy, ``An adi-pe approach for modeling radio
  transmission loss in tunnels,'' \emph{IEEE Transactions on Antennas and
  Propagation}, vol.~57, no.~6, pp. 1759--1770, 2009.

\bibitem{paradacomparative}
D.~A. Parada, C.~G. Rego, C.~G. Batista, and G.~L. Ramos, ``A comparative study
  between sspe methods and a ho-fdtd algorithm for em propagation over lossy
  terrains,'' in \emph{EuCAP 2018}, 2018.

\bibitem{popov2000modeling}
A.~V. Popov and N.~Y. Zhu, ``Modeling radio wave propagation in tunnels with a
  vectorial parabolic equation,'' \emph{IEEE Transactions on Antennas and
  Propagation}, vol.~48, no.~9, pp. 1403--1412, 2000.

\bibitem{sevgi2005matlab}
L.~Sevgi, {\c{C}}.~Ulu{\i}{\c{s}}{\i}k, and F.~Akleman, ``A matlab-based
  two-dimensional parabolic equation radiowave propagation package,''
  \emph{Antenna and Propagation Magazine}, pp. 164--175, 2005.

\bibitem{shen2012comparisons}
Y.~SHEN, L.~ZHANG, D.~LIU, Y.~WU, L.~MU, and R.~C. HUNTSINGER, ``Comparisons of
  ray-tracing and parabolic equation methods for the large-scale complex
  electromagnetic environment simulations,'' \emph{International Journal of
  Modeling, Simulation, and Scientific Computing}, vol.~3, no.~02, p. 1240005,
  2012.

\bibitem{taj2016new}
I.~A. Taj-Eddin, M.~Afifi, M.~Korashy, D.~Hamdy, M.~Nasser, and S.~Derbaz, ``A
  new compression technique for surveillance videos: Evaluation using new
  dataset,'' in \emph{Digital Information and Communication Technology and its
  Applications (DICTAP), 2016 Sixth International Conference on}.\hskip 1em
  plus 0.5em minus 0.4em\relax IEEE, 2016, pp. 159--164.

\bibitem{wang2016parabolic}
D.-D. Wang, X.-L. Xi, Y.-R. Pu, J.-F. Liu, and L.-L. Zhou, ``Parabolic equation
  method for loran-c asf prediction over irregular terrain,'' \emph{IEEE
  Antennas and Wireless Propagation Letters}, vol.~15, pp. 734--737, 2016.

\bibitem{wang2012propagation}
K.~Wang and Y.~Long, ``Propagation modeling over irregular terrain by the
  improved two-way parabolic equation method,'' \emph{IEEE Transactions on
  Antennas and Propagation}, vol.~60, no.~9, pp. 4467--4471, 2012.

\bibitem{yun2015ray}
Z.~Yun and M.~F. Iskander, ``Ray tracing for radio propagation modeling:
  Principles and applications,'' \emph{IEEE Access}, vol.~3, pp. 1089--1100,
  2015.

\bibitem{zhang2015efficient}
X.~Zhang, N.~Sood, J.~Siu, and C.~D. Sarris, ``Efficient propagation modeling
  in railway environments using a hybrid vector parabolic equation/ray-tracing
  method,'' in \emph{Antennas and Propagation \& USNC/URSI National Radio
  Science Meeting, 2015 IEEE International Symposium on}.\hskip 1em plus 0.5em
  minus 0.4em\relax IEEE, 2015, pp. 1680--1681.

\bibitem{zhang2014hybrid}
X.~Zhang, N.~Sood, J.~Siu, and C.~D. Sarris, ``A hybrid vector parabolic
  equation/ray-tracing propagation modeling technique for rail transportation
  systems,'' in \emph{Radio Science Meeting (Joint with AP-S Symposium), 2014
  USNC-URSI}.\hskip 1em plus 0.5em minus 0.4em\relax IEEE, 2014, pp. 123--123.

\bibitem{zhang2016hybrid}
X.~Zhang, N.~Sood, J.~K. Siu, and C.~D. Sarris, ``A hybrid ray-tracing/vector
  parabolic equation method for propagation modeling in train communication
  channels,'' \emph{IEEE Transactions on Antennas and Propagation}, vol.~64,
  no.~5, pp. 1840--1849, 2016.

\bibitem{zhao2013new}
X.-f. Zhao, S.-x. Huang, and L.-c. Kang, ``New method to solve electromagnetic
  parabolic equation,'' \emph{Applied Mathematics and Mechanics}, vol.~34,
  no.~11, pp. 1373--1382, 2013.

\end{thebibliography}

\end{spacing}


%
%

\end{document}